# Local Realistic Model of Bell Theorem Experiment and Alternative Model of Quantum Measurement

Paul J. Werbos[1]
ECCS Division, National Science Foundation

## Abstract

A previous paper demonstrated that two local, realistic models based on Markov Random Fields (MRF) across space-time do replicate the correct, tested predictions of quantum mechanics for simple Bell's Theorem experiments. This paper demonstrates a third such model, MRF3, which is more plausible physically, making contact with the physics of polarizers. It shows how to translate that type of model into an equivalent quantum mechanical model – which requires an augmentation to the usual measurement operator now used to represent polarizers in such experiments. The validity of that augmentation might be testable in experiments involving three or four photons, with relevance to quantum computing and communication.

## 1. Introduction

A previous paper [1] discussed the important practical challenge of trying to build models which can correctly predict the behavior of systems embodying quantum entanglement, as best we can subject to the constraints of classical computing in three dimensions. Since Bell's Theorem experiments provide a clear, well-defined example of quantum entanglement, I focused on that example. I showed that two relatively simple lumped parameter models correctly predict the outcomes of the basic experiment.

Those two models were based on the simple mathematics of Markov Random Fields (MRF), which are well-known in computer science. In the language of J.S. Bell [2], these were both local realistic models – but they do not violate his theorems, because his theorems only rule out local realistic models which proceed by making feedforward calculations running forwards in time. These were iterative models, based on recurrent calculations. I will refer to those models here as MRF1 (the more complicated model) and MRF2 (a simpler model, exploiting an additional variable representing the direction of time which a photon is moving in).

Section 2 of this paper presents a new model of the same Bell experiment, MRF3, which makes better contact with what we know about the physics of polarizers [3] and with the modeling techniques used in quantum optics [4,5,6]. The main result of that section is that MRF3, like MRF1 and MRF2, correctly replicates the predictions of quantum theory for the basic Bell's Theorem experiment.

Section 3 of this paper discusses what happens if we translate MRF3 in a more general way into an equivalent, time-forwards quantum mechanical model, based on density matrices rather than wave functions as in modern quantum optics [5,6]. In traditional quantum mechanics, the ideal polarizer is represented as a simple projection operator, a measurement operator. MRF3 suggests a modification to that operator. Thus, even though MRF3 and quantum theory agree for the case of Bell's Theorem experiments, they might well disagree for more complex experiments with entanglement, perhaps for systems of three or four photons. Section 4 will make a few brief comments about the status of such experiments; so far as I know, we do not yet have a decisive experiment in hand showing which form of the measurement operator fits better, the original Copenhagen form or the modified form suggested in section 3. It should be noted that this model results in a "positive P" state at all times in the experiment [6].

---

[1] The views expressed here are those of the author, not those of his employer; however, as work produced on government time, it is in the "government public domain." This allows unlimited reproduction, subject to a legal requirement to keep the document together, including this footnote and authorship.

MRF3 is still just an approximate model, like the models given by Carmichael himself, even though they fit a huge body of empirical data with greater accuracy than many simple "exact" models. Nevertheless, the usual quantum mechanical calculations for Bell's Theorem experiments are also done at a high level of abstraction [1,2,7]. The formulation of the measurement operator describing what polarizers do has implications well beyond the simple case discussed in section 2.

In [1], I asked whether the simpler model (MRF2) might have a higher probability of truth than MRF1, which appeared messier and more complicated. In machine learning, when two models both fit empirical data, but one is simpler, the simpler model usually has a higher probability of truth. Occam's Razor is a very fundamental principle in machine learning [8]. Here, however, we are trying to model a complex system with complex emergent behavior. The underlying Boltzmann picture [9] is the simplest theory of all of these, but it attributes at least some small nonzero probability to a wide variety of possible trajectories, even wider than the variety analyzed in MRF1. MRF3 is more complicated than MRF1 in some ways, but closer to the underlying physics, and suitable for generalization to a wider variety of other experiments.

The most important question for future research here is whether the equivalence between MRF3 and quantum mechanics extends to more systems than just the Bell's Theorem experiments. Section 4 will suggest some ideas for such future research, focusing on the next logical set of examples, those involving three or four entangled photons.

## 2. Specification and Results of the MRF3 Model

### 2.1 Review of MRF1 and the Experiment to Be Predicted

This section will show that the new MRF3 model can correctly predict the outcome of the most simple, basic form of Bell's Theorem experiment, illustrated in Figure 1.

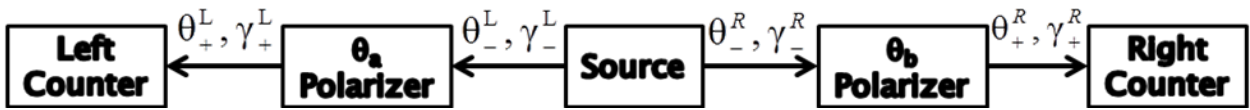

Figure 1. Core Structure and Notation for the First Bell's Theorem Experiments

Figure 1 is taken directly from [1]. It is an abstract version of the design shown in Figure 2 of Clauser and Shimony [8], the most definitive primary source on the original experiments performed by Clauser, Holt and others. Figure 1 also shows the eight scenario variables used in the MRF1 model, which was called "the transparent model" in [1]. In this experiment, the Source produces two entangled photons, one of which goes to the left channel (L), and one of which goes to the right channel (R). The four binary variables γ represent the presence or absence of a photon traveling along the initial or later part of one of these channels. The four continuous variables θ represent the angle of polarization of such a photon, when it is present. A "scenario" is a set of values for all eight variables.

In the usual language of Markov Random Fields, figure 1 defines an undirected graph. The eight random variables along the edges of that graph define a possible state or scenario of the system defined by the graph.

The goal of the MRF1 model was simply to predict the correct rate of coincidences, as detected by the counters, as a fraction of those scenarios in which two entangled photons were in fact produced. In other words, the goal is to predict the frequency or probability of scenarios in which $\gamma_+^L = \gamma_+^R = 1$, conditional upon $\gamma_-^L = \gamma_-^R = 1$. The observed and predicted coincidence rates are a function of $\theta_a$ and $\theta_b$, the angles of polarization chosen by the experimenter. In these experiments, these were always chosen to be linear polarizers. Quantum mechanics predicts that the rate of two-photon detections, as a fraction of the rate of production of two entangled photons, is:

$$R_2(\theta_a-\theta_b)/R_0 = \tfrac{1}{2}\cos^2(\theta_a-\theta_b) \qquad (1)$$

Of course, when $\theta_a$ is orthogonal to $\theta_b$, the rate is zero – a prediction which seems very mysterious to many people, and cannot be reconciled with classical ways of thinking. This is the essence of why this is an important experiment.

See [1] for a review of the derivation of (1) in quantum mechanics, and for some additional details not relevant to our goals here.

The MRF1 model is much simpler than the usual model from quantum mechanics, even though it replicates the same correct predictions, as shown in [1]. Like MRF3, it is a discrete Markov Random Field (MRF) model. This means that it predicts the probability of any scenario X in two steps:

$$P^*(X) = p_1^*(X)p_2^*(X)p_3^*(X)...p_n^*(X) \qquad (2)$$
$$Pr(X) = P^*(X)/Z , \qquad (3)$$

where the "partition function" Z is simply a constant used to make the probabilities add up to 1. More precisely, Z is the sum or integral of $P^*(X)$ over all possible scenarios X. I will refer to the quantities $P^*$ and $p^*$ as "relative probabilities." To complete the specification of any discrete MRF model, we must enumerate the n nodes in the graph, and model the endogenous probability function $p_n^*(X)$ characterizing each node. (The endogenous probability functions are often called "feature functions" in computer science.)

For the case of MRF1, the five nodes were simply the five objects illustrated in Figure 1 – the source, the two detectors and the two polarizers. The endogenous probability model for the polarizer was a very abstract input-output model, as is the usual measurement operator M representing the polarizer in Copenhagen quantum mechanics. For MRF3, we will provide a small amount of additional detail, still providing a simple high-level model, but making contact with the actual physics of what goes on in one kind of polarizer used in these experiments.

MRF1 was basically the simplest, highest level model possible able to replicate the correct coincidence rate, other than MRF2, which is not physically plausible. Even so, MRF1 made use of a model of the detector which may not seem like the simplest possible model

$$p^*(\gamma_+, \theta_+) = 1 - \gamma_+ + \alpha\gamma_+ d\theta_+ \qquad (4)$$

This relative probability is not normalized to one; that is not required, because equation 3 takes care of normalization. More important, this model asserts a high probability that no photon will be detected, and a probability on the order of $\alpha$, a small number, for detecting a photon of any polarization $\theta_+$. It also assumes a uniform distribution for possible polarization angles, which is what we expect for a detector equally able to detect any polarization. Implicitly, we assume that if a photon is actually there (that $\gamma_+=1$), the detector will in fact detect it.

In classical thinking, we would make $\alpha$ a large number, such as $1/2\pi$. That is how we normally think about what a detector does, as we march forwards in time. However, the Bell's Theorems [2] show that it is impossible to replicate the correct observed result of equation (1) if we pick probabilities which implement the usual time-forwards classical picture. Thus in MRF1 and in MRF3, we pick endogenous probabilities which are symmetric with respect to time – Endogenous Time-Symmetric Probabilities (ETSP) – for all objects *except* the Source, which receives free energy flowing forward from the past., providing the power to this experiment.

The time-symmetric view of photon emission and absorption seems counterintuitive at first; however, the original paper on the photoelectric effect by Einstein showed how a time-symmetric representation does work. Roughly speaking, the $\alpha$ here is on the order of Einstein's A coefficient, a function of the temperature of the system. Section 4 will begin to address questions about the underlying physics at a deeper level; here, however, our goal is simple to specify a new MRF model, and show that it works.

## 2. 2. Structure of the MRF3 Model and Picture of the Experiment

The MRF3 model, like MRF1, is still a relatively abstract and high-level model, but it adds the details needed to represent the essence of what actually happens in a detector and in one of the types of polarizers used in these experiments.

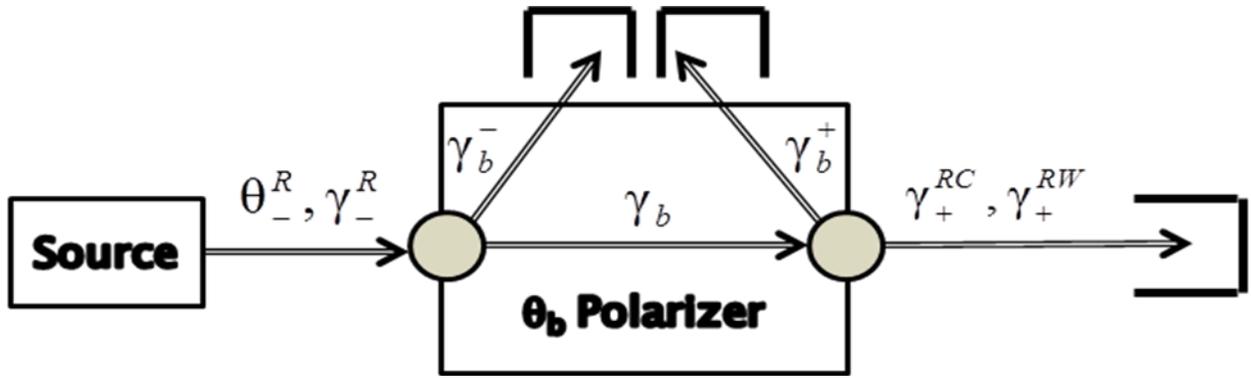

Figure 2. MRF3 picture of the right channel of the experiment

Figure 2 illustrates the picture for the right channel of the experiment. In effect, Figure 2 zooms in on the right half of Figure 1, providing more detail. MRF3 also assumes the mirror image of this picture, to model the left channel of the experiment.

Here, instead of two nodes to the right of the source (as in MRF1), we assume five nodes: the incoming and outgoing surfaces of the polarizer denoted by circles, the usual detector on the right, and two "hidden detectors" inside the polarizer which absorb the light from internal reflection. This gives a reasonable high-level picture of what actually happens in a calcite crystal polarizer [3], the choice made in the first of the Bell's Theorem experiments [8]. (It is also consistent with the gross behavior of the dichroic polarizer, used in many of the other experiments, discussed in less detail in Lipson [8].)

Even today, studies of objects like calcite crystals rely heavily on Maxwell's Laws with parameters like the index of refraction estimated to reflect properties of the crystal. This is all just an approximation, since we know that a complete description of the crystal would model every atom in the crystal; however, it is a useful and important approximation, with a huge body of empirical work to support it. It is less extreme of an approximation than simply modeling the crystal as a simple projection operator.

From that work using Maxwell's Laws [8], we know that the light adapts very quickly to the boundary conditions represented by the incoming and outgoing surfaces. Here, as in Lipson [8], we will approximate the situation by assuming an instantaneous effect at the two boundaries. Within the crystal, in this experiment, we will assume that there either does or does not exist a photon inside the crystal, propagating in the direction detectable in this experiment, with the linear polarization which can escape from the boundaries of the crystal; thus, there is just one binary variable here, $\gamma_b$, representing the presence or absence of such a photon traveling through the crystal. There are two additional binary variables, $\gamma_b^-$ and $\gamma_b^+$, representing the existence of photons with polarizations such that they are captured by internal reflection, and end up being absorbed by one or the other of two hidden internal virtual detectors.

The representation of the photon counter on the right is also more realistic here than in Figure 1, and more consistent with what we know from quantum field theory. We assume that the detector on the right, like a two-level atom, can only absorb photons which are circularly polarized. Thus $\gamma_+^{RC}$ is a binary variable representing the existence of a photon with a clockwise circular polarization (C), and $\gamma_+^{RW}$ represents the existence of a photon with a counterclockwise or "widdershins" circular polarization (W).

A more complete model would also allow for possibilities like circular polarization coming from the source. We leave it as an exercise (a fairly easy one) to see why this results in additional terms of higher order in the key statistical parameters $\alpha$ and $\beta$, such they do not contribute effects measurable at the levels of precision applied to these experiments to date. Likewise, modeling the internal absorbers in more detail does not really change the outcomes to be shown here.

## 2.3 Models of the Endogenous Probabilities (ETSP) For the Five Types of Object

For the source, I use the same ETSP model as in MRF1 [1], but am explicit that we are only interested in events where two entangled photons are produced:

$$p^*(\theta_-^L, \gamma_-^L, \theta_-^R, \gamma_-^R) = \delta(\theta_-^L - \theta_-^R)\gamma_-^L \gamma_-^R \tag{5}$$

For the two external detectors, the new model used in MRF3 is:

$$p^*(\gamma^C, \gamma^W) = 1 \quad \text{when } \gamma^C = \gamma^W = 0 \tag{6}$$
$$= \alpha \quad \text{when } \gamma^C = 1 \text{ or } \gamma^W = 1$$

As discussed below, $\alpha$ mirrors the Einstein A coefficients; intuitively, in MRF3, it represents the thermodynamic cost or difficulty of absorbing a circularly polarized photon. We also assume that the counter will report a detection when and only when $\gamma^C = 1$ or $\gamma^W = 1$.

For the four hidden internal detectors, we use a model similar to that used in MRF1:

$$p^*(\gamma, \theta) = 1 - \gamma + 2\alpha\beta\gamma d\theta, \tag{7}$$

where $\theta$ is the polarization of the photon coming into the detector (if it exists). The small parameter $\beta$ reflects the thermodynamic cost, in the crystal, of converting a linearly polarized photon into a circularly polarized one, which can then be absorbed with a cost of $\alpha$. $\beta$ and $\alpha$ are just parameters in this abstract model, but are actually the outcome of thermodynamic processes below the level of resolution of MRF3. The factor of 2 represents the presence of two options for circular polarization, parallel to what we model in more detail for the external detector.

For the crystal surface closest to the source, we use:

$$p^*(\theta_-, \gamma_b, \gamma_b^-) = \delta(\theta_- - \theta)\gamma_b + \delta(\theta_- - \theta - \pi/2)\gamma_b^- + \beta(\cos^2(\theta_- - \theta)\gamma_b + \sin^2(\theta_- - \theta)\gamma_b^-) \tag{8}$$

except $= 0$ for $\gamma_b = \gamma_b^-$

where $\theta$ is the angle which the polarizer is tuned to (either $\theta_a$ or $\theta_b$) and where $\beta$ and $\alpha$ are as above. This is the relative probability for the specific case where $\gamma_-=1$, the case when there is an incoming photon which is linearly polarized. Notice that there is no thermodynamic cost effect, $\beta$, for simply passing through a linearly polarized photon already aligned with one of the two principal allowed polarizations in the crystal, for photons moving in the direction determined by this experiment.

Finally, for the surface furthest from the crystal we use:

$$\begin{aligned} p^*(\gamma_b, \gamma^C, \gamma^W) &= 0 \quad \text{for } \gamma^C = \gamma^W = 1 \\ &= 0 \quad \text{for } \gamma_b = 1 \text{ and } \gamma^C = \gamma^W = 1 \\ &= \beta \quad \text{for } \gamma_b = 1 \text{ and } \gamma^C = 1 \text{ or } \gamma^W = 1 \quad (9) \\ &= \beta \quad \text{for } \gamma_b = 0 \text{ and } \gamma^C = 1 \text{ or } \gamma^W = 1 \\ &= 1 \quad \text{for } \gamma_b = 0 \text{ and } \gamma^C = \gamma^W = 0 \end{aligned}$$

Notice that equation 9 does not explicitly include $\gamma_b^-$. We do not need to include $\gamma_b^-$ explicitly, because we assume that $\gamma_b^- = 1$ when and only when $\gamma_b = 0$ and $\gamma^C = 1$ or $\gamma^W = 1$

Because equations 9 and 10 look quite different, it is not immediately obvious that this model of the polarizer is consistent with the goal of time-symmetric representation [9] for objects which do not inject a flow of free energy into the experiment. That is because the experimental setup gives low priority to anything but circular polarization after the polarizer, and linear before, according to the model; it is easy enough to unify equations 9 and 10 into a common time-symmetric model, allowing for possibilities like circular photons coming out of the source to the polarizer; however, since such scenarios have low probability, there is no need to include that detail here. Our goal here is to present the simplest MRF model possible, which makes more concrete contact with the actual physics than MRF1 did.

MRF3 actually uses 10 variables to define a possible scenario -- $\theta_-^R$, $\gamma_b$, $\gamma_b^-$, $\gamma_+^{RC}$, $\gamma_+^{RW}$, $\theta_-^L$, $\gamma_a$, $\gamma_a^-$, $\gamma_+^{LC}$ and $\gamma_+^{LW}$. It assumes that $\gamma_-^L = \gamma_-^+ = 1$ for all scenarios of interest here, and that we do not need to consider cases where $\gamma_b^+ = 1$ or $\gamma_a^+ = 1$. That compares with six variables actually used in MRF1. In that respect, MRF3 is a more complicated model. However, equations 8 and 9 together are still much simpler and more natural than the polarizer model in MRF1. Also, only two of the scenario variables in MRF3 ($\theta_-^L$ and $\theta_-^R$) are continuous variables, giving only one effective continuous degree of freedom (since the probability is zero for any scenario in which they do not equal each other). In MRF4, there were four continuous variables, and three continuous degrees of freedom. Thus in some respects, MRF3 is actually the simpler model, and far more plausible from a physical point of view.

## 2.4. Predictions of the Model

The main result of this section is that the MRF3 model does replicate the correct probabilities given in equation 1, as a function of $\theta_a$ and $\theta_b$, in the limit as $\alpha$ and $\beta$ go to zero. We will mainly focus on the case where $\theta_a$ does not equal $\theta_b$ or $\theta_b + \pi/2$ exactly.

As in [1], we verify this by simply calculating what the predictions of the model are. Here we will take the same general approach as in [1], but will organize the calculation in a more abstract way, so as to avoid the explicit tabulation of $2^8 = 256$ possible scenarios for each possible value of $\theta_-$.

Let us begin by defining the logical variable "D" so as to mean "there was a double photon count observed in this scenario for the outcome of the experiment." Let us define:

$$P_+^*(\theta_-) = \sum_{X \in D} p_1^*(X) p_2^*(X) ... p_{10}^*(X) \qquad (10)$$

and

$$P_-^*(\theta_-) = \sum_{X \in (\text{not } D)} p_1^*(X) p_2^*(X) ... p_{10}^*(X) \qquad (11)$$

where X refers to a set of values for all the scenario variables other than $\theta_-^L$, $\theta_-^R$, $\gamma_-^L$ and $\gamma_+^R$, and where we assume ten objects in the model, five on the right and five on the left, other than the source itself. Recall that $R_0$ in equation 1 refers to the total rate of scenarios in which two entangled photons are produced, and $R_2$ refers to the rate in which two entangled photons are produced and in which D is true [1,8]; thus to predict that ratio, we need only calculate relative probabilities for scenarios in which $\gamma_-^L = \gamma_-^R = 1$ and in which $\theta_-^L = \theta_-^R = \theta_-$, where we still need to consider all possible values of the continuous scenario variable $\theta_-$. From equations 2 and 3, it is easy to see that MRF3 predicts the ratio $R_2/R_0$ to be the following probability conditional upon the emission of two entangled photons at the source:

$$\Pr(D) = (1/Z) \int P_+^*(\theta_-) d\theta_- : \qquad (12)$$

where:

$$Z = \int \left( P_+^*(\theta_-) + P_-^*(\theta_-) \right) d\theta_- \qquad (13)$$

Next let us defined $D_L$ and $D_R$ as logical variables representing a photon detection at the external counter on the far left and the far right, respectively. Because the scenario variables on the left channel and the scenario variables on the right channel do not interact with each other, in this model, after $\theta_-$ is specified, it is convenient to define $X_L$ as the set of scenario variables for the left channel, $X_R$ for the right. We can then define:

$$P_{+L}^*(\theta_-) = \sum_{X_L \in D_L} p_1^*(X_L) p_2^*(X_L) ... p_5^*(X_L) \qquad (14)$$

$$P_{-L}^*(\theta_-) = \sum_{X_L \in (\text{not } D_L)} p_1^*(X_L) p_2^*(X_L) ... p_5^*(X_L) \qquad (15)$$

and likewise for the right channel. It is straightforward to see

$$P_+^*(\theta_-) = P_{+L}^*(\theta_-) P_{+R}^*(\theta_-) \qquad (14)$$

$$P_+^*(\theta_-) + P_-^*(\theta_-) = \left( P_{+L}^*(\theta_-) + P_{-L}^*(\theta_-) \right) \left( P_{+R}^*(\theta_-) + P_{-R}^*(\theta_-) \right) \qquad (15)$$

To calculate $P_{+R}^*(\theta_-)$ and $P_{-R}^*(\theta_-)$, consider the picture in Figure 2.
For any given value of $\theta_-$ and scenario $X_R$ for the right channel, there are just two scenarios of nonzero probability allowed in this model, with $D_R$ true:

(1) $\gamma_b = 1$, and $\gamma_+^{RC} = 1$, implying $\gamma_b^- = \gamma_-^{RW} = 0$
(2) $\gamma_b = 1$, and $\gamma_+^{RW} = 1$, implying $\gamma_b^- = \gamma_-^{RC} = 0$

The relative probability along the right channel for scenario (1) is just the product of the partial probabilities for the incoming surface (from equation 8), for the outgoing surface (equations 9) and the external detector (equation 6):

$$(\delta(\theta_- - \theta_b) + \beta \cos^2(\theta_- - \theta_b))(\beta)(\alpha)$$

Note that "$\theta$" in equation 8 referred to the angle of a polarizer in general, but now we are considering the right channel, whose polarization is set by the experimenter to $\theta_b$. We get the exact same expression for scenario (2), resulting in the sum:

$$P_{+R}^*(\theta_-) = 2(\delta(\theta_- - \theta_b) + \beta \cos^2(\theta_- - \theta_b))(\beta)(\alpha) \qquad (16)$$

Likewise, we have just one right-channel scenario in which $D_R$ is false:

(3) $\gamma_b^- = 1$, implying $\gamma_b = \gamma^{RC} = \gamma^{RW} = 0$

The relative probability for this scenario is the product of the relative probability for $\gamma_b^-$ at the incoming surface of the polarizer (from equation 8) and the relative probability for the internal hidden detector (equation 7):

$$P_{-R}^*(\theta_-) = (\delta(\theta_- - \theta_b - \pi/2) + \beta \sin^2(\theta_- - \theta_b))(2\beta\alpha) \qquad (17)$$

Combining equations 16 and 17, we easily see that:

$$P_{+R}^*(\theta_-) + P_{-R}^*(\theta_-) = 2\beta\alpha(\delta(\theta_- - \theta_b) + \delta(\theta_- - \theta_b - \pi/2)) + 2\beta^2\alpha \qquad (18)$$

The reported Bell's Theorem experiments generally do not address experiments where $\theta_a$ equals $\theta_b$ exactly, with or without a ninety degree rotation, which is really just a limit of the cases where they are not equal. Thus when we substitute equation 18 and its left-channel equivalent into equations 15, and integrate over $\theta_-$, as in equation 13, we simply get:

$$Z = (2\pi)4\beta^4\alpha^2 \qquad (19)$$

Likewise, equation 16 and its left-hand equivalent give us:

$$P_{+R}^*(\theta_-)P_{+L}^*(\theta_-) = 4\beta^2\alpha^2(\delta(\theta_- - \theta_a)\delta(\theta_- - \theta_b) + \beta\delta(\theta_- - \theta_a)\cos^2(\theta_- - \theta_b) \\ + \beta\delta(\theta_- - \theta_b)\cos^2(\theta_- - \theta_a) + \beta^2\cos^2(\theta_- - \theta_a)\cos^2(\theta_- - \theta_b)) \qquad (20)$$

When we integrate this over θ₋, as called for in equation 12, the first term drops out, because the two delta functions multiplied together will always give zero when $\theta_a$ is not equal to $\theta_b$, as we are assuming here. The last term is of higher order in β, and also drops out the limit as β goes to zero. Performing the integration, and substituting into equation 12, we then deduce what the prediction of MRF3 is for this experiment:

$$\Pr(D) = 8\beta^3\alpha^2\cos^2(\theta_a - \theta_b)/Z = \cos^2(\theta_a - \theta_b) \qquad (21)$$

which matches equation 1 as claimed. QED.

For the situation where $\theta_a$ equals $\theta_b$ exactly, the predicted probability is dominated by just two scenarios – the scenario where $\theta_- = \theta_b = \theta_a$ and D=1, and the case where $\theta_- = \theta_b+\pi/2=\theta_a+\pi/2$ and D=0. Since these are predicted to have equal probability, Pr(D) is predicted to be ½, which also fits equation 1. Likewise, when $\theta_a = \theta_b+\pi/2$, the probabilities are dominated by two scenarios, in both of which D=0, consistent with equation 1. Because these special case predictions involve the products of Dirac delta functions, they are mathematically rigorous only if we interpret equation 5 as a limit of a regularized version, using the function g to be discussed in the following section.

## 3. The Quantum Mechanical Equivalent to MRF3

### 3.1. Translation From MRF Models to Quantum Mechanics In General

The first step in translating from an MRF model to quantum mechanics is to translate the MRF model into a more traditional Markovian kind of model, marching forwards in time. This can be done by considering the evolution of the function Pr(X(t),t), where X(t) is the set of all scenario variables in play at time t, conditional upon all the decisions made up to time t.

For example, here there are four variables in play -- $\gamma_-^L$, $\gamma_-^R$, $\theta_-^L$ and $\theta_-^R$ – at time $t_-+\varepsilon$, where $t_-$ is the start of this round of the experiment, when two photons come out of the source. We assume $\gamma_-^L=\gamma_-^R=1$ in that case; $\Pr(X(t_-+\varepsilon),t_-+\varepsilon) = \delta(\theta_-^L-\theta_-^R)d\theta_-^R$, from equation 5. That probability distribution remains in effect at later times, until the time when a photon first hits a polarizer, on the left or on the right. At each time when an event like that happens, we perform a Bayesian convolution of Pr(X(t-ε),t-ε) with the endogenous probability distribution of the object/event, to derive Pr(X(t+ε),t+ε). That sequence of n Bayesian convolutions yields equations 2 and 3, when we reach the nth and final object. This computational model is technically "nonlocal" in the sense of Bell's Theorems [2,7], because of the way in which Bayesian convolution works. (It is nonlocal in the same way that the ordinary many-worlds version of quantum field theory is nonlocal; many worlds quantum theory is also realistic, but is consistent with the Bell's Theorem experiments, because of this limited degree of nonlocality.)

The second step, in general, is to map the probability distribution Pr(X(t),t) at any time t into the corresponding density matrix. This can be done by use of the Glauber-Sudarshan P mapping [4,5,9], which can be used to map any probability distribution for classical states into a corresponding density matrix.

## 3.2 Review of the Traditional Polarizer/Measurement Model

Consider what happens when a photon is emitted from some source, travels to a polarizer, and then – if not absorbed – travels from the polarizer to a human eye which acts as a detector/observer. According to the oldest version of quantum mechanics, the photon is in a "mixed state" between the polarizer and the human eye. It is like a Schrodinger cat, which is neither alive nor dead but in a mixed state until the precise moment when a human observer, with metaphysical observer status, looks at it. Here, I will consider the case of a calcite crystal without (or before) the internal reflection, such that the photon which emerges from the initial contact is in a mixed state, between the two possible polarizations in the crystal ($\gamma_b$ and $\gamma_b^-$, in effect).

Modern quantum optics does not require that the detector be a human eyeball. Mandel has explained this by saying "we do not actually need to observe… it is enough that we threaten to observe." More concretely, it is assumed that the polarizer itself acts like a kind of measurement operator. When a wave function $|\psi\rangle$ hits the initial surface of a calcite crystal, the outgoing wave is in a mixed state of $|\theta_b\rangle$ and $|\theta^b+\pi/2\rangle$. Thus the measurement operation or "quantum jump" [5] at this point maps any pure state $|\psi\rangle$ by the linear mapping to a mixed state:

$$M: \quad \rho(t-\varepsilon) = |\psi\rangle\langle\psi| \;\rightarrow\; c_b |\theta_b\rangle\langle\theta_b| + c_b^- |\theta_b+\pi/2\rangle\langle\theta_b+\pi/2| \qquad (22)$$

where $c_b=(\langle\theta_b|\rho(t-\varepsilon)|\theta_b\rangle)$ and $c_b^- = 1-c_b$. Since this is a linear mapping, we can decompose *any* incoming density matrix $\rho$ into pure states, and derive the result of this simple measurement operation or "quantum jump." Note that M is an example of what Carmichael [4,5] calls a "superoperator,' a linear mapping from the space of density matrices or operators to the space of density matrices or operators. Intuitively, wave functions are like vectors; density matrices and ordinary operators are like matrices; and superoperators are like fourth order tensors.

## 3.3. The Polarizer/Measurement Model of MRF3

Section 2.4 shows us that the predictions of MRF3 are essentially based on equation 8, the model of what happens when a photon first encounters a calcite crystal. Equation 8 plays essentially the same role as that of the usual measurement operator in traditional quantum mechanical calculations [1,2,7]. Aside from the factor of $\beta$, the right-hand side of equation 8 is essentially the same as the traditional operator.

Equation 8 implies a new polarizer model for incoming density matrices (with a slight generalization). For any incoming density matrix $\rho$, assume a decomposition into pure states of linear polarization $\theta_*$ plus a sum $\rho_0$ of a residual component (e.g. pure states of circular polarization):

$$\rho = \rho_0 + \int c_\theta \, |\theta_*\rangle\langle\theta_*| \, d\theta_* \qquad (23)$$

Then define the linear superoperator $M_0^*$ by:

$$M_0^*: \rho_0 \rightarrow \beta M(\rho_0) \tag{24}$$

where M is defined as in equation 22, and

$$M_0^*: |\theta_-\rangle\langle\theta_-| \rightarrow g(\theta-\theta_-)M(|\theta_-\rangle\langle\theta_-|) \tag{25},$$

where $g(\theta-\theta_-)$ is a function which may be $\delta(\theta-\theta_-)+\delta(\theta-\theta_--\pi/2)+\beta$, or a close approximation. The overall new polarizer model, $M^*$, is the normalized version of $M_0^*$:

$$M^*: \rho \rightarrow M_0^*(\rho)/\text{Tr}(M_0^*(\rho)) \tag{26}$$

Of course, this superoperator is nonlinear. In practical calculations, it would often make sense to follow Carmichael's quantum trajectory approach [5], by simply tabulating the different discrete possibilities ($\theta_-=\theta$, $\theta_-=\theta-\pi/2$, other) and running them forwards in parallel. In theory, the function g could be represented as an equivalent linear operator in the P representation, to make $M^*$ linear, but at least for now it is hard to see that much need to enforce linearity here. In the future, if we use the MRF approach just to handle what Carmichael calls the "quantum jumps," and his other methods for the flow of events between those jumps, it will be possible to be more explicit about issues like the known timing constraints for interference effects in three-photon and four-photon experiments.

This leads to the obvious question: given that correct predictions for the Bell experiment have been obtained both from MRF3 and from the more traditional version of quantum mechanics, would the same be true for more complicated experiments involving polarizers, such as the GHZ family of three-photon and 4 photon experiments, which have served as a pathway to some important work on quantum computing? At present, we simply do not know; that would appear to be the most important question for further research related to MRF3.

## **4. Open Questions for Relevant 3 and 4 Photon Experiments**

The logical next follow-on to MRF3 would be to evaluate its ability to predict experiments involving three or four entangled photons.

There has been extensive research into entangled sets of three or photons, initially inspired by the seminal "GHZ" paper [10]. That paper proposed a triple interferometry experiment, which would entail the generation of three photons, entangled in the sense that:

$$\theta_a + \theta_b + \theta_c = 0 \tag{27}$$

Reck, in Zeilinger's group, reported in 1996 that no source had been found as yet bright enough to perform this experiment [11]. In 1998, Timothy Keller proposed a way to meet this challenge, building this equation [12], which resulted in a paper with his advisers Rubin and Shi, and with Wu [13]. Also in 1998, Bouwmeester et al from Zeilinger's group reported that they had been able to produce GHZ states experimentally [14]; in 2000, they reported that they had achieved predictions consistent with quantum mechanics in that system, ruling out local, causal realistic models of physics in much the same way as the earlier Bell's Theorem experiments did [15]. Building on that work, Zeilinger's group has moved into substantial new work extending into ever more practical designs for use in quantum computing.

However, since MRF3 is not a "causal' model, in the sense of the original Bell's theorems, and since we have seen that it can fit the Bell's theorem results, these results are not directly transferable to the choice between M and M* (or a direct use of MRF methods) to the case of three or four photons. One line of possible work here is to extend the explicit MRF modeling to experiments which have already been done, to verify its viability there. More interesting, perhaps, would be the design and performance of new experiments explicitly designed to decide between alternative theories here.

As an example, consider the special kind of "triphoton" experiment depicted in figure 3:

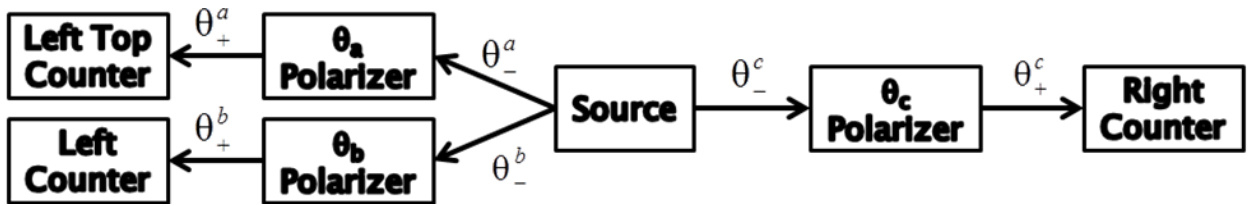

Figure 3. A possible triphoton experiment

In the MRF3 model, it is clear that it does not matter which of the three photons reaches its polarizer first. The probability calculations are made on a graph, in a way which does not depend on which of the three photons reaches a polarizer first, or on which one exits a polarizer first. But in the traditional Copenhagen view of measurement, a measurement operator acts immediately in the wave function; in the modern version of the Copenhagen view, the density matrix for the system of three photons exiting the polarizers is predicted to be:

$$\rho_+ = M(\theta_3)M(\theta_2)M(\theta_1)\rho_- \qquad (28)$$

where $\rho_-$ is the density matrix generated by the source, and the subscripts 1, 2 and 3 refer to the order in which the photons reach their polarizers. These measurement superoperators are essentially the same as those which govern a single beam of light propagating through three polarizers in sequence; when there are three polarizers in such a sequence, moving one polarizer just one millimeter ahead or behind another can have a drastic effect on the outcome, because these linear superoperators do not commute with each other. If an experiment and source can be found such that a slight change in the timing of photon arrival in Figure 3 changes the traditional predictions of quantum mechanics, that experiment would provide a way to discriminate between MRF3 and the usual measurement model. In a sense, this would be like another round of the Einstein-Podolosky-Rosen challenge, where the triumph of quantum mechanics might well lead to another useful form of what Kimble calls "quantum weirdness."

On the other hand, if such an experiment cannot be devised, it is possible that MRF3 and the usual measurement model will continue to yield identical predictions for cases involving three and four photons, and more – giving us a way to calculate predictions for more complicated quantum systems in three dimensions. Either way, it is an important question for future research.